\documentclass{article}

\usepackage[usenames,dvipsnames,svgnames,table]{xcolor}
\usepackage[breaklinks=true,hyperfootnotes=false,colorlinks,bookmarks=false,citecolor=MidnightBlue,linkcolor=BrickRed,urlcolor=MidnightBlue]{hyperref}
\usepackage{nips13submit_e,times}
\usepackage[round]{natbib}
\usepackage{graphicx}
\usepackage{paralist}
\usepackage{amsmath}
\usepackage{booktabs}

\graphicspath{{figures/}}

\newcommand{\eg}{\emph{e.g.},}

\newcommand{\ie}{\emph{i.e.},}

\newcommand{\etc}{\emph{etc.}}
\newcommand{\vs}{\emph{vs.}}

\DeclareUrlCommand\emailurl{}

\newcommand{\eaname}[3]{\href{mailto:#3}{#1}\thanks{#2}\\\email{#3}}
\newcommand{\efname}[3]{\href{mailto:#3}{#1}\footnotemark[#2]\\\email{#3}}

\title{The Compositional Nature of Verb and Argument Representations in the
  Human Brain}

\author{
\eaname{Andrei Barbu}{School of Electrical and Computer
  Engineering, Purdue University, West Lafayette IN 47907-2035}
{andrei@0xab.com}
\And
\efname{N. Siddharth}{1}{snarayan@purdue.edu}
\And
\eaname{Caiming Xiong}{Department of Computer Science and
  Engineering, SUNY Buffalo, Buffalo NY 14260-2500}
{cxiong@buffalo.edu}
\And
\efname{Jason J. Corso}{2}{jcorso@buffalo.edu}
\And
\eaname{Christiane D. Fellbaum}{Department of Computer
  Science, Princeton University, Princeton NJ 08540-5233}
{fellbaum@princeton.edu}
\And
\eaname{Catherine Hanson}{Department of Psychology and Rutgers Brain Imaging
  Center, Rutgers University, Newark NJ 07102}
{cat@psychology.rutgers.edu}
\And
\efname{Stephen Jos\'{e} Hanson}{4}{jose@psychology.rutgers.edu}
\And
\eaname{S\'{e}bastien H\'{e}lie}{Psychological Sciences, Purdue University, West
  Lafayette IN 47907}
{shelie@purdue.edu}
\And
\eaname{Evguenia Malaia}{Southwest Center for Mind, Brain, and Education,
  University of Texas at Arlington, Arlington TX 76019}
{malaia@uta.edu}
\And
\eaname{Barak A. Pearlmutter}{Hamilton Institute \& Dept Computer Sci, National
  University of Ireland Maynooth, Co.\ Kildare, Ireland}
{barak@cs.nuim.ie}
\And
\efname{Jeffrey Mark Siskind}{1}{qobi@purdue.edu}
\And
\efname{Thomas Michael Talavage}{1}{tmt@purdue.edu}
\And
\eaname{Ronnie B. Wilbur}{Department of Speech, Language, and Hearing Sciences
  and Linguistics Program, Purdue University, West Lafayette IN 47907}
{wilbur@purdue.edu}
}

\nipsfinalcopy %

\begin{document}

\maketitle

\begin{abstract}
  How does the human brain represent simple compositions of objects, actors,
  and actions?
  We had subjects view action sequence videos during neuroimaging (fMRI)
  sessions and identified lexical descriptions of those videos by decoding
  (SVM) the brain representations based only on their fMRI activation patterns.
  As a precursor to this result, we had demonstrated that we could reliably
  and with high probability decode action labels corresponding to one of
  six action videos (\emph{dig}, \emph{walk}, \etc), again while subjects
  viewed the action sequence during scanning (fMRI).
  This result was replicated at two different brain imaging sites with common
  protocols but different subjects, showing common brain areas, including areas
  known for episodic memory (PHG, MTL, high level visual pathways, \etc,
  \ie\ the `what' and `where' systems, and TPJ, \ie\ `theory of mind').
  Given these results, we were also able to successfully show a key aspect of
  language compositionality based on simultaneous decoding of object class and
  actor identity.
  Finally, combining these novel steps in `brain reading' allowed us to
  accurately estimate brain representations supporting compositional decoding
  of a complex event composed of an actor, a verb, a direction, and an object.
\end{abstract}

\section{Introduction}
\label{sec:introduction}

The compositional nature of thought is taken for granted by many in the
cognitive-science and artificial-intelligence communities.
For example, in computer vision, representations for nouns, such as those
used for object detection, are independent of representations for verbs, such
as those used for event recognition.
Humans need not employ compositional representations; indeed, many argue that
such representations may be doomed to failure in AI systems
\citep{brooks1991intelligence}.
This is because concepts like \emph{verb} or even \emph{object} are human
constructs; there is debate as to how they arise from percepts
\citep{smith1996origin}.
Recent advances in brain-imaging techniques enable exploration of the
compositional nature of thought.
To that end, subjects underwent functional magnetic resonance imaging (fMRI)
during which they were exposed to stimuli which evoke complex brain activity
which was decoded, piece by piece.
The video stimuli depicted events described by entire sentences composed of a
\emph{verb}, an \emph{object}, an \emph{actor} and a \emph{location} or
\emph{direction} of motion.
By decoding complex brain activity into its constituent parts, we show evidence
for the neural basis of the compositionality of verb and argument
representations.

Recent work on decoding brain activity corresponding to nouns has
recovered object identity from nouns presented as image and orthographic
stimuli.
\citet{hanson2009} perform classification on still images of two object
classes: faces and houses, and achieve an accuracy above 93\% on a
one-out-of-two classification task.
\citet{connolly2012} perform classification on still images of objects, two
instances of each of three classes: bugs, birds, and primates, and achieve an
accuracy between 60\% and 98\% on a one-out-of-two within-class classification
task and an accuracy between 90\% and 98\% on a one-out-of-three between-class
classification task.
\citet{just2010} perform classification on orthographically presented nouns, 5
exemplars from each of 12 classes, achieving a mean rank accuracy of 72.4\% on
a one-out-of-60 classification task, both within and between subjects.
\citet{pereira2012} incorporate semantic priors and achieve a mean accuracy
of 13.2\% on a one-out-of-12 classification task and 1.94\% on a one-out-of-60
classification task when attempting to recover the object being observed.
\citet{miyawaki2008} recover the position of an object in the field of view by
recovering low resolution images from the visual cortex.
Object classification from video stimuli has not been previously demonstrated.

Recent work on decoding brain activity corresponding to verbs has primarily
been concerned with identifying active brain regions.
\citet{kable2006} present the brain regions which attempt to distinguish
between the different agents of actions and between the different kinds of
actions they perform.
\citet{kemmerer2008} analyze the regions of interest (ROI) of brain activity
associated with orthographic presentation of twenty different verbs in each of
five different verb classes.
\citet{kemmerer2010} analyze the brain activity associated with verbs in
terms of the motor components of event structure and attempt to localize the
ROIs of such motor components.
While prior work analyzes regions which are activated when subjects are
presented verbs as stimuli, we recover the content of the resulting brain
activity by classifying the verb from brain scans.

Recent work demonstrates the ability to decode the actor of an event using
personality traits.
\citet{Hassabis-etal-2013a} demonstrate the ability to recover the identity of
an imagined actor from that actor's personality.
Subjects are informed of the two distinguishing binary personality traits of
four actors.
During fMRI, they are presented sentences orthographically which describe an
actor performing an action.
The subjects are asked to imagine this scenario with this actor and to rate
whether the actions of the actor accurately reflect the personality of that
actor.
The resulting brain activation corresponding to these two binary personality
traits is used to recover the identity of the actor.
No prior work has recovered the identity of an actor without relying on that
actor's personality.
In the work presented here, the personality of the actor has no bearing on the
actions being performed.

In this paper, two new experiments are presented.
In Experiment~1, subjects are shown videos and asked to think of verbs that
characterize those videos.
Their brains are imaged via fMRI and measured neural activation is
decoded to recover the verb that the subjects are thinking about.
Decoding is done by means of a support vector machine (SVM) trained on brain
scans of those same verbs.
We know of no other work that decodes brain activity corresponding to verbs.
We show early evidence that the regions identified by this decoding process are
not intimately tied to a particular subject \emph{via} an additional analysis
that trains on one subject and tests on another.
In Experiment~2, subjects are shown videos and asked to think of complex
sentences composed of multiple components that characterize those videos.
We show a novel ability to decode brain activity corresponding to multiple
objects: the identity of an actor and the identity of an object.
We decode the identity of an actor without relying on the personality traits of
that actor.
We know of no other work which recovers an entire sentence composed of multiple
constituents.
We find evidence that suggests underlying neural representations of mental
states are independent and compose into sentences largely without modifying one
another.

\section{Compositionality}
\label{sec:compositionality}

We discuss a particular kind of compositionality as it applies to sentence
structure: objects fill argument positions in predicates that combine to form
the meaning of a sentence.
\citet{pylkkanen2011grounding} reviews work which attempts to show this kind of
compositionality using a task called \emph{complement coercion}.
Subjects in this task are presented with sentences whose meaning is richer
than their syntax.
For example, the sentence \emph{The boy finished the pizza} is understood as
meaning that the pizza was eaten, even though the verb \emph{eat} does not
appear anywhere in the sentence \citep{pustejovsky1995}.
The presence of \emph{pizza}, belonging to the category \emph{food}, coerces
the interpretation of \emph{finish} as \emph{finish eating}.
By contrast, \emph{He finished the newspaper} induces the interpretation
\emph{finish reading}.
Because the syntactic complexity in this prior experiment was held constant, the
assumption is that coercion is a purely semantic meaning-adding function
application, with little consequence for the syntax.
The participants completed this task, and brain activity was measured using
magnetoencephalography (MEG).
The results show activity related to coercion in the anterior midline field.
This result suggests an initial localization for at least some function
application, but it is difficult to use MEG to distinguish whether this
activity is read from the ventromedial prefrontal cortex or the anterior
cingulate cortex.
Earlier work on the representation of objects and actions in the brain also
indicates that these representations may be independent.

\paragraph{Representing objects in the brain}
Objects are static entities that can be represented by a (mostly) static
neural representation.
For example, the~3D representation of a soda can will look the same in many
different contexts, and the appearance of the soda can is not unfolding in
time.
It is generally believed that the lexicon of object concepts is represented in
the medial temporal lobe while different areas of the temporal lobe may be
combinatoric in constructing object types \citep{hanson2004combinatorial}
although there may be modal areas associated with different representational
functions.
For example, lesion data suggests that the temporal pole is associated with
naming people, the inferior temporal cortex is associated with naming
animals, and the anterior lateral occipital regions are associated with naming
tools.
In addition, some regions involved in object representation are
modality specific.
For example, spoken-word processing involves the superior temporal lobe (part
of the auditory associative cortex; \citealp{binder2000human}) while reading
words representing objects activates occipito-temporal regions because of the
visual processing \citep{puce1996differential}.
Specifically, auditory word processing involves a stream of information
starting in Heschl's gyri that is transferred to the superior temporal gyrus.
Once the superior temporal gyrus has been reached, the modality of stimulus
presentation is no longer relevant.
In contrast, the initial processing for written words starts in the occipital
lobe (V1 and V2), and moves on to occipito-temporal regions specialized in
identifying orthographic units.
The information then moves rostrally to the temporal lobe proper, where
modality of presentation is no longer relevant \citep{binder2000human}.

\paragraph{Representing actions in the brain}
Unlike objects, verbs are dynamic entities that unfold in time.
For instance, observing someone pick up a ball takes time as the person's
movement unfolds.
Evidence reviewed in \citet{coello2012motor} suggests that action verbs
activate both semantic units in the temporal cortex and a motor network.
The motor network includes the premotor areas (including the supplementary
motor area), the primary motor cortex, and the posterior parietal cortex.
Some researchers went as far as suggesting that the well-known ventral/dorsal
distinction in the visual pathways corresponds to a semantic (ventral) and
action (dorsal) distinction.
Representation of action may involve `mirror neurons' that have been shown in
macaque to respond jointly in perception/action tasks, where the similarity of
the self action is to the perceived action of an observed individual.

\section{Approach}
\label{sec:approach}

All experiments reported follow the same procedure and are analyzed using the
same methods and classifiers.
Videos are shown to subjects who are asked to think about some aspect(s) of the
video while whole-brain fMRI scans are acquired every two seconds.
Because fMRI acquisition times are slow, roughly equal to the length of the
video stimuli, a single brain volume that corresponds to the brain activation
induced by that video stimulus is classified to recover the features that the
subjects were asked to think about.
Multiple runs separated by several minutes of rest, where no data is acquired,
are performed per subject.

\subsection{fMRI procedures}

Imaging performed at Purdue University used a 3T GE Signa HDx scanner
(Waukesha, Wisconsin) with a Nova Medical (Wilmington, Massachusetts) 16
channel brain array to collect whole-brain volumes via a gradient-echo EPI
sequence with 2000ms TR, 22ms TE, 200mm$\times$200mm FOV, and 77$^{\circ}$ flip
angle.
We acquired 35 axial slices with a 3.000mm slice thickness using a 64$\times$64
acquisition matrix resulting in 3.125mm$\times$3.125mm$\times$3.000mm voxels.

Imaging performed at St.~James Hospital in Dublin, Ireland, used a 3T
Phillips Achieva scanner (Best, The Netherlands) using a gradient-echo EPI
sequence with 2000ms TR and 240mm$\times$240mm FOV.\@
We acquired 37 axial slices with a 3.550mm slice thickness using an
80$\times$80 acquisition matrix resulting in
3.000mm$\times$3.000mm$\times$3.550mm voxels.

\subsection{fMRI processing}

Data was acquired in runs, with between three and eight runs per subject per
experiment, and each axis of variation of each experiment was counterbalanced
within each run.
fMRI scans were processed using AFNI \citep{cox1996afni} to skull-strip each
volume, motion correct and detrend each run, and align each subject's runs to
each other.
Voxels within a run were z-scored, subtracting the mean value of that voxel for
the run and dividing by its variance.
Because each brain volume has very high dimension, between 143,360 and
236,800 voxels, we eliminate voxels by computing a per-voxel Fisher score on
our training set and keeping the 5,000 highest-scoring voxels.
The Fisher score of a voxel $v$ for a classification task with $C$~classes
where each class~$c$ has $n_c$ examples is computed as
\begin{equation}
  \frac{\displaystyle\sum_{c=1}^C n_c(\mu_{c,v}-\mu)^2}
       {\displaystyle\sum_{c=1}^C n_c \sigma_{c,v}^2}
\end{equation}
where $\mu_{c,v}$ and $\sigma_{c,v}$ are the per-class per-voxel means and
variances and $\mu$ is the mean for the entire brain volume.
A linear SVM classifies the selected voxels.

One run was taken as the test set and the remaining runs were taken as the
training set.
The third brain volume after the onset of each stimulus was taken along with
the class of the stimulus to train an SVM.\@
This lag of three brain volumes is required because fMRI does not measure
neural activation but instead measures the flow of oxygenated blood, the
blood-oxygen-level-dependent (BOLD) signal, which correlates with increased
neural activation.
It takes roughly five to six seconds for this signal to peak which puts the
peak in the third volume after the stimulus presentation.
Cross validation was performed by choosing each of the different runs as the
test set.

To understand our results and to demonstrate that they are not classifying
noise or irrelevant features, we perform an analysis to understand the brain
regions that are relevant to each experiment.
We determine these regions by two methods.
First we employ a spatial searchlight \citep{kriegeskorte2006information} which
slides a small sphere across the entire brain volume and repeats the above
analysis keeping only the voxels inside that sphere.
We use a sphere of radius three voxels, densely place its center at every
voxel, and do not perform any dimensionality reduction on the remaining voxels.
We then perform an eight-fold cross validation as described above for each
position of the sphere.
For Experiment~1 we also back-project the SVM coefficients onto the anatomical
scans---the higher the absolute value of the coefficient the more that voxel
contributes to the classification performance of the SVM---and use a classifier
with a different metric, $w(i)^2$, as described by \citet{hanson2009}.

\section{Experiment 1: Verb Representation}
\label{sec:experiment1}

\begin{figure}[t]
  \begin{center}
    \setlength{\tabcolsep}{1.5pt}
    \begin{tabular}{@{}ccc@{\hspace{10pt}}ccc@{}}
      \includegraphics[width=0.15\textwidth]{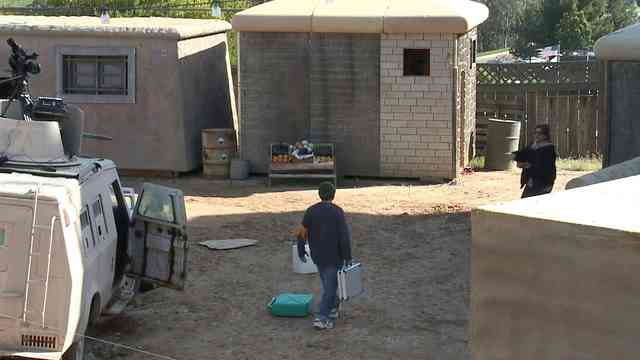}&
      \includegraphics[width=0.15\textwidth]{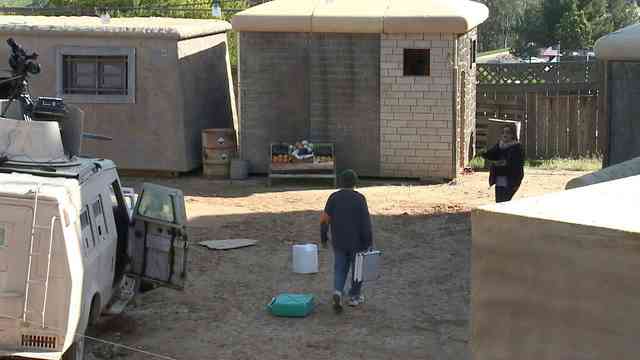}&
      \includegraphics[width=0.15\textwidth]{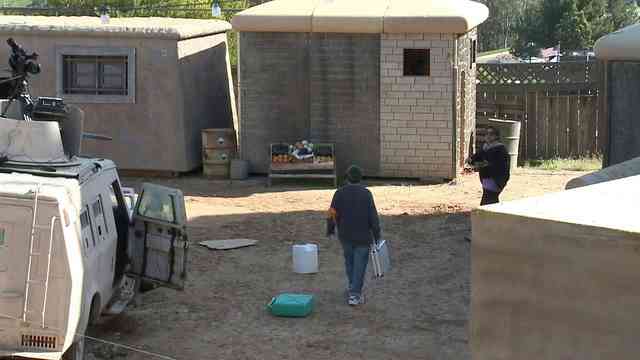}&
      \includegraphics[width=0.15\textwidth]{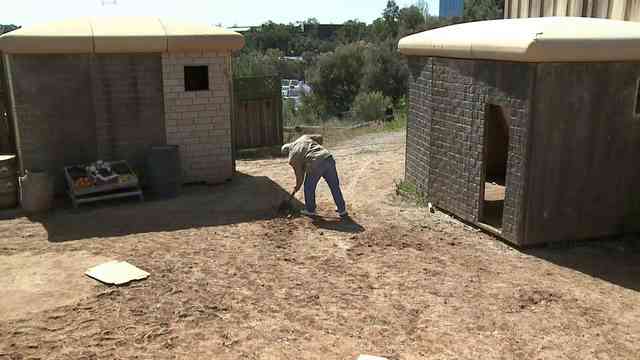}&
      \includegraphics[width=0.15\textwidth]{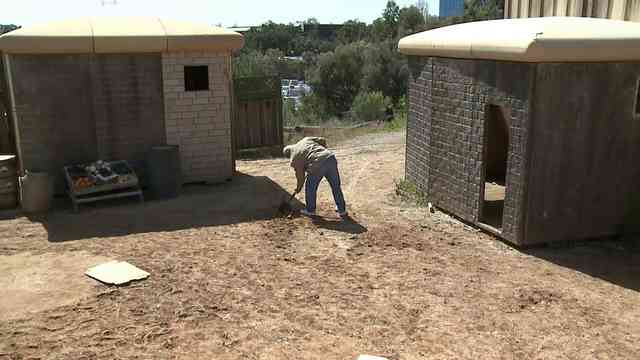}&
      \includegraphics[width=0.15\textwidth]{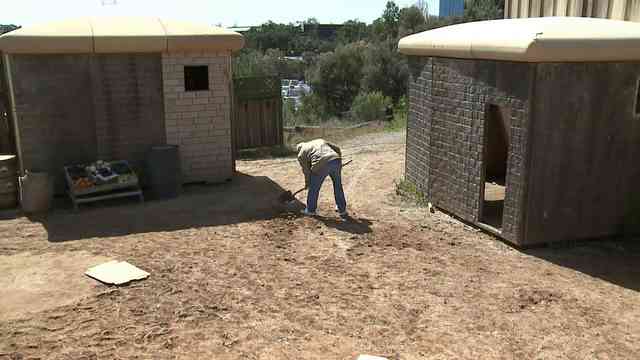}\\[-0.8ex]
      \multicolumn{3}{c@{\hspace{10pt}}}{\emph{carry}}&
      \multicolumn{3}{c}{\emph{dig}}\\[0.5ex]
      \includegraphics[width=0.15\textwidth]{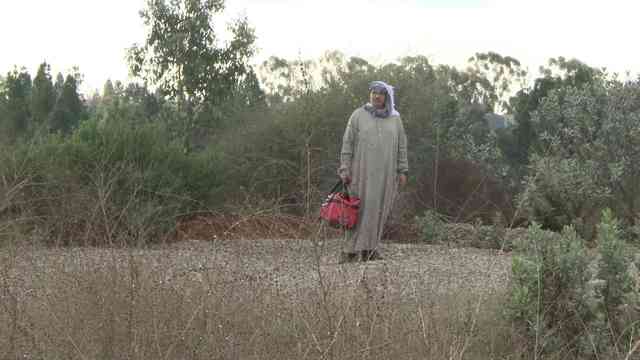}&
      \includegraphics[width=0.15\textwidth]{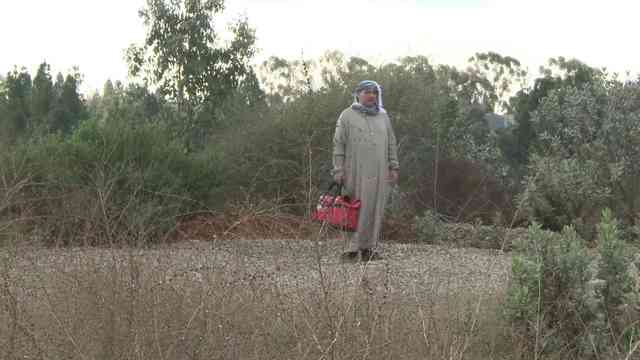}&
      \includegraphics[width=0.15\textwidth]{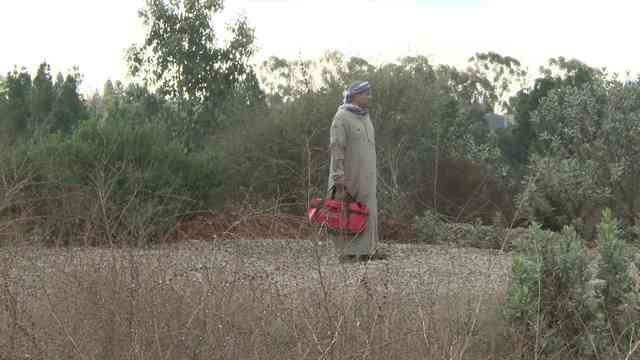}&
      \includegraphics[width=0.15\textwidth]{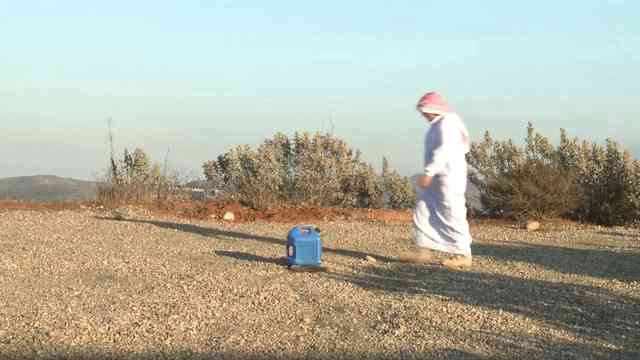}&
      \includegraphics[width=0.15\textwidth]{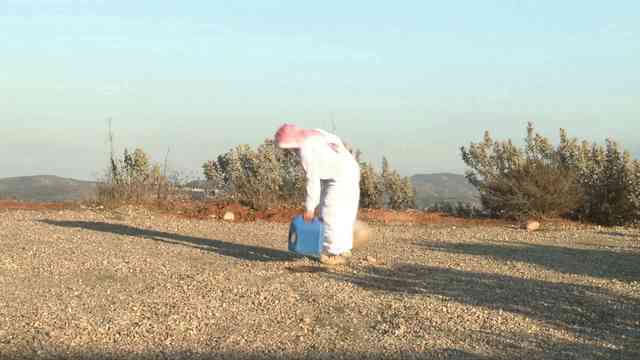}&
      \includegraphics[width=0.15\textwidth]{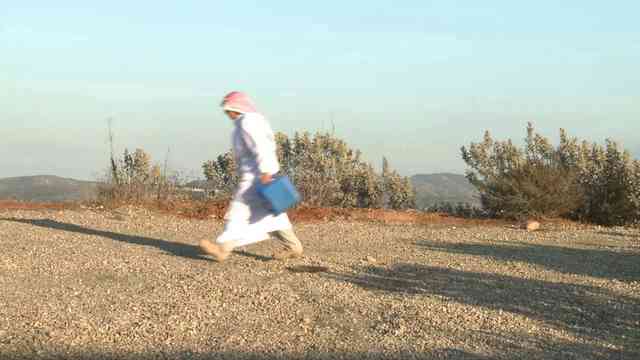}\\[-0.8ex]
      \multicolumn{3}{c@{\hspace{10pt}}}{\emph{hold}}&
      \multicolumn{3}{c}{\emph{pick up}}\\[0.5ex]
      \includegraphics[width=0.15\textwidth]{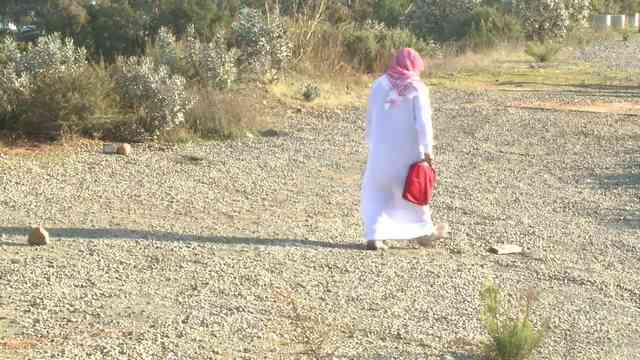}&
      \includegraphics[width=0.15\textwidth]{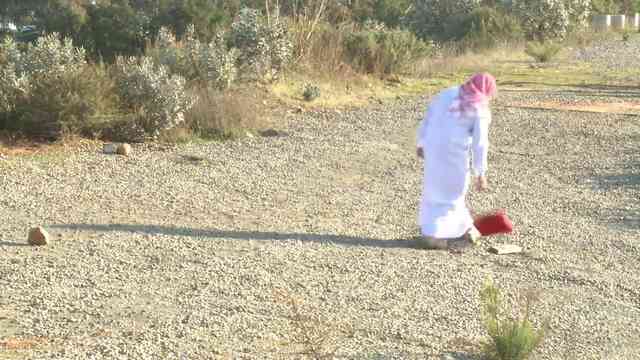}&
      \includegraphics[width=0.15\textwidth]{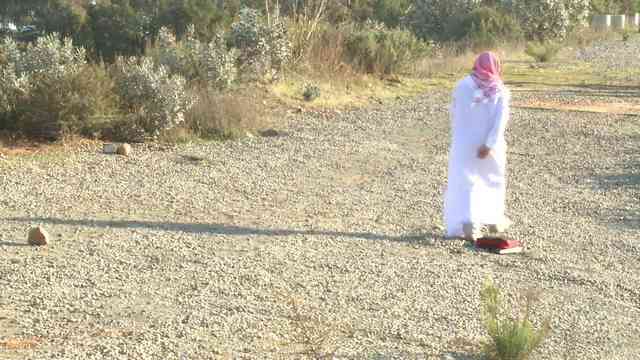}&
      \includegraphics[width=0.15\textwidth]{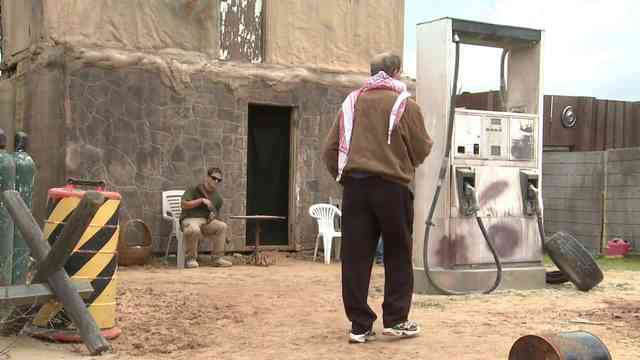}&
      \includegraphics[width=0.15\textwidth]{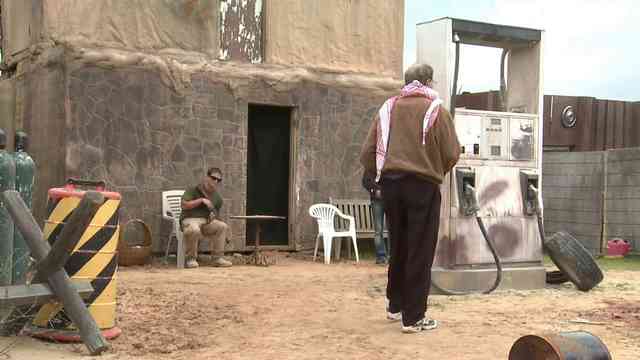}&
      \includegraphics[width=0.15\textwidth]{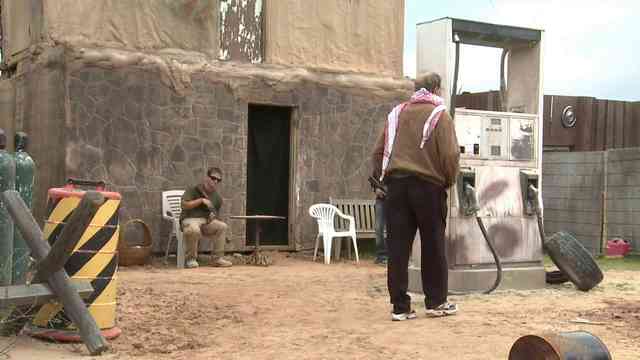}\\[-0.8ex]
      \multicolumn{3}{c@{\hspace{10pt}}}{\emph{put down}}&
      \multicolumn{3}{c}{\emph{walk}}
    \end{tabular}
  \end{center}
  \caption{Key frames from sample stimuli for each of the six verbs in
    Experiment~1.
  Example stimulus videos are included in the supplementary material.}
  \label{fig:ME-Y2}
\end{figure}

We conducted an experiment to evaluate the ability to identify brain activity
corresponding to verbs denoting actions.
Subjects are shown video clips of humans interacting with objects and are told
to think of the verb being enacted, but otherwise have no task.
The subjects were shown clips depicting each of these verbs prior to the
experiment and were instructed about the intended meaning of each verb.
One difficulty with such an experiment is that there is disagreement between
human subjects as to whether a verb occurred in a video or not.
To overcome this difficulty, we asked five humans to annotate the
DARPA Mind's Eye year 2 video corpus with the extent of every verb.
From this corpus, we chose video clips where at least two out of the five
annotators agreed on the depiction.
We selected between twenty seven and thirty 2.5s video clips depicting each of
six different verbs (\emph{carry}, \emph{dig}, \emph{hold}, \emph{pick up},
\emph{put down}, and \emph{walk}).
Key frames from one clip for each of the six verbs are shown in
Fig.~\ref{fig:ME-Y2}.
Despite multiple annotators agreeing on whether a video depicts a verb,
the task of classifying each clip remains very difficult for human subjects
as it is easy to confuse similar verbs such as \emph{carry} and \emph{hold}.
We address this problem by presenting, in rapid succession, pairs of video
clips which depict the same verb and asking the subjects to think about the
verb that would best describe both videos.

We employed a rapid event-related design similar to that of \citet{just2010}.
We presented pairs of 2.5s video clips at 12fps, depicting the same
verb, separated by 0.5s blanking and followed by an average of 4.5s (minimum
2.5s) fixation.
While the video clips within each pair depicted the same verb, the clips across
pairs within a run depicted different verbs, randomly counterbalanced.
Each run comprised 48 stimulus presentations spanning 254 captured brain volumes
and ended with 24s of fixation.
Eight runs for each of subjects~1 through~3 were collected at Purdue University.
Three runs for subject~4 and four runs for subject~5 were collected at
St.~James Hospital.

\begin{figure}
  \begin{center}
      \includegraphics[width=0.45\textwidth]{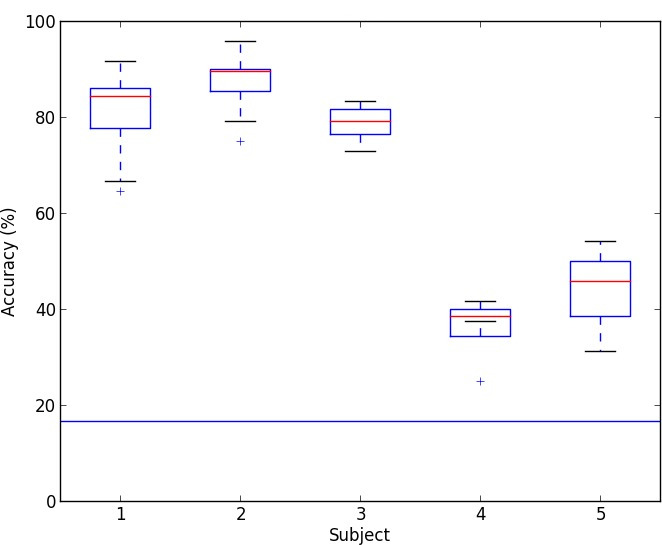}
      \hfill
      \includegraphics[width=0.45\textwidth]{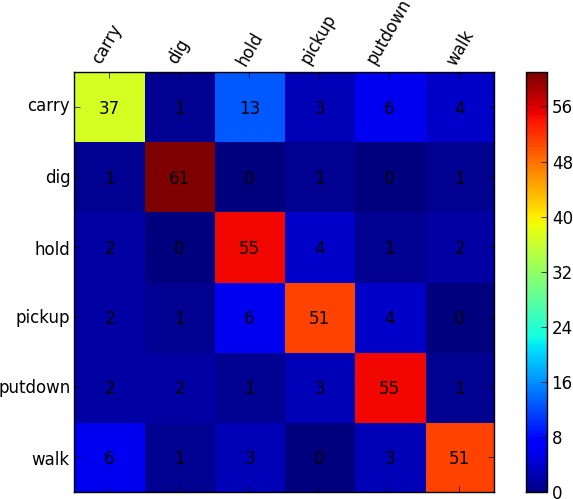}
  \end{center}
  \caption{Results for Experiment~1.
    (left)~Per-subject classification accuracy on 1-out-of-6 verb classes
    averaged across class and fold.
    Horizontal line indicates chance performance, 16.66\%.
    (right)~Corresponding confusion matrix averaged across subject and fold
    is mostly diagonal, with the highest numbers of errors
    being made distinguishing \emph{hold} and \emph{carry}, two ambiguous
    stimuli.}
  \label{fig:ME-Y2-results}
\end{figure}

We performed an eight-fold cross validation (fewer for subjects~4 and~5) for a
six-way classification task, where runs constituted folds.
The results are presented in Fig.~\ref{fig:ME-Y2-results}.
The per-subject accuracies, averaged across class and fold, were: 80.73\%,
87.24\%, 78.91\%, 35.94\%, and 43.75\% (chance 16.66\%).
Note that the last two were trained on fewer runs than the first three.
This demonstrates the ability to recover the verb that the subjects were
thinking about.
The robustness of this result is enhanced by the fact that it was replicated on
two different fMRI scanners at different locations run by different
experimenters.

To evaluate whether the brain regions used for classification generalize across
subjects, we performed an additional analysis on the data for subjects~1 and~2.
One run out of the eight was selected as the test set and the data for one of
the two subjects was classified.
The training set consisted of all seven other runs for the subject whose
data does not appear in the test set.
The test was performed on the run omitted from the training set, even though it
was gathered from a different subject, to preclude the possibility that the
same stimulus sequence appeared in both the training and test sets.
We performed cross validation by varying which subject contributes the test
data and which subject contributes the training data, and within each of these
folds we varied which of the eight runs is the test set.
These two cross validations yielded accuracies of 33.59\%
(subject~1$\mapsto$subject~2) and 41.41\% (subject~2$\mapsto$subject~1),
averaged across class and fold, where chance again is 16.66\%.

\begin{figure}[t]
  \centering
  \begin{tabular}{c}
    \includegraphics[width=\textwidth]{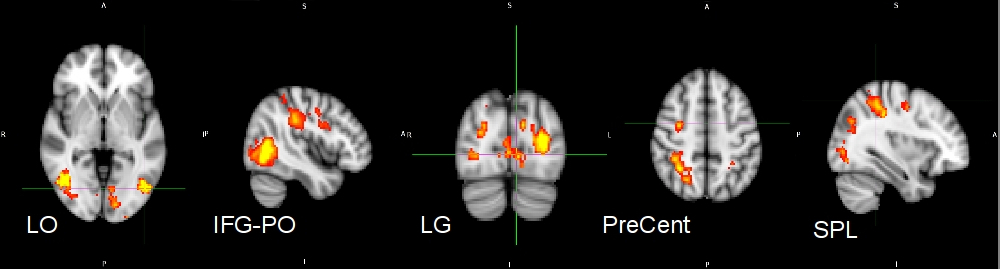}\\
    \includegraphics[width=\textwidth]{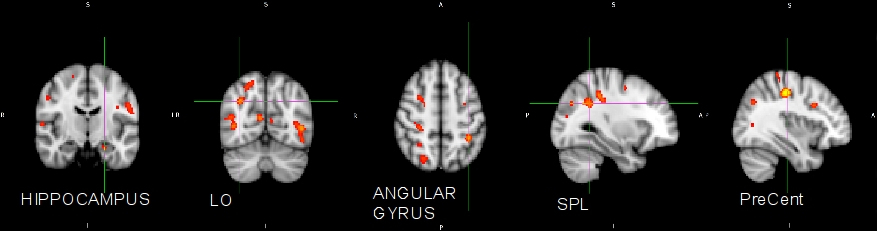}
  \end{tabular}
  \caption{(top)~Searchlight analysis for Experiment~1 indicating the
    classification accuracy of different brain regions on the anatomical scans
    from subject~1, averaged across stimulus, class, and run.
    (bottom)~A similar analysis using a $w(i)^2$ metric.}
  \label{fig:ME-Y2-searchlight}
\end{figure}

To locate regions of the brain used in the previous analysis, we used a
spatial-searchlight linear-SVM method on subject~1.
We use the accuracy to determine the sensitivity of each voxel and threshold
upward to less then 5\% of the cross-validation measures.
These measures are overlaid and (2-stage) registered to MNI152 2mm
anatomicals shown in Fig.~\ref{fig:ME-Y2-searchlight}(top).
Notable are visual-pathway areas (lateral occipital-LO, lingual gyrus-LG, and
fusiform gyrus) as well as prefrontal areas (inferior frontal gyrus, middle
frontal gyrus, and cingulate) and areas consistent with the `mirror system'
\citep{arbib2006action} and the so-called `theory of mind' (pre-central gyrus,
angular gyrus-AG, and superior parietal lobule-SPL) areas
\citep{dronkers2004lesion,turken2011neural}.
Fig.~\ref{fig:ME-Y2-searchlight}(bottom) shows the decoded ROIs from a similar
SVM classifier with a different metric, $w(i)^2$ \citep{hanson2009}, showing
similar brain areas but, due to higher sensitivity, also indicates sub-cortical
regions (hippocampal) associated with encoding processes not seen with the
cross-validation accuracy metric.
As argued in Section~\ref{sec:compositionality}, lateral-occipital areas are
involved in visual processing specifically related to language, and the
fusiform gyrus is a hetero-modal area that could hold abstract representations
of the elements contained in the videos (\eg\ semantics).
This data brings initial support for the hypothesis that concepts have both
modality-specific and abstract representations.
Hence, the elements used by the SVM to classify the videos are also
neuroscientifically meaningful.

\section{Experiment 2: Argument Representation}
\label{sec:experiment2}

\begin{figure}
  \begin{center}
    \setlength{\tabcolsep}{1pt}
    \begin{tabular}{@{}ccc@{\hspace{10pt}}ccc@{\hspace{10pt}}ccc@{}}
      \includegraphics[width=0.1\textwidth]{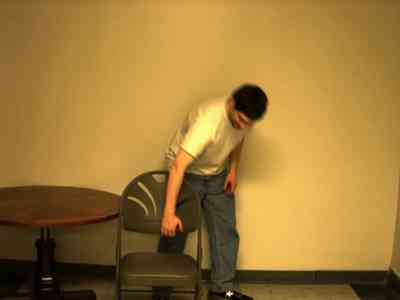}&
      \includegraphics[width=0.1\textwidth]{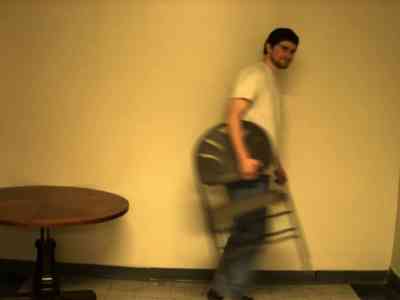}&
      \includegraphics[width=0.1\textwidth]{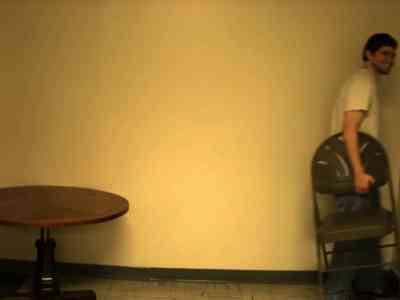}&
      \includegraphics[width=0.1\textwidth]{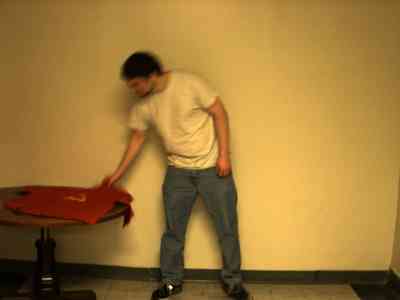}&
      \includegraphics[width=0.1\textwidth]{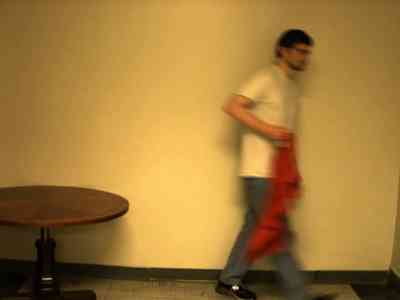}&
      \includegraphics[width=0.1\textwidth]{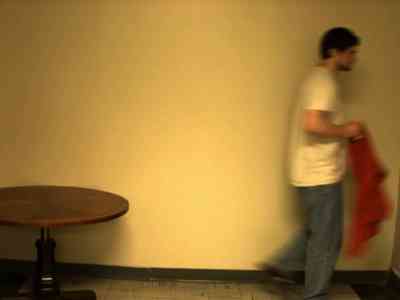}&
      \includegraphics[width=0.1\textwidth]{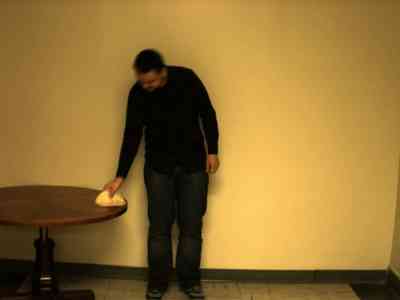}&
      \includegraphics[width=0.1\textwidth]{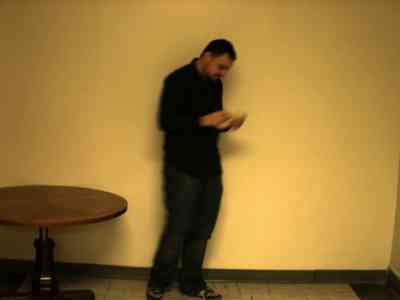}&
      \includegraphics[width=0.1\textwidth]{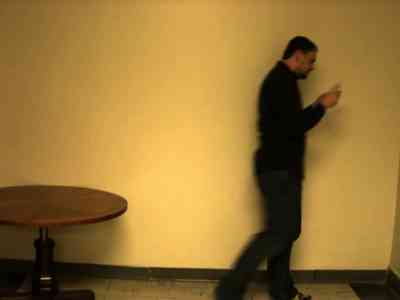}\\[-0.8ex]
      \multicolumn{3}{c@{\hspace{10pt}}}{\emph{carry chair}}&
      \multicolumn{3}{c@{\hspace{10pt}}}{\emph{carry shirt}}&
      \multicolumn{3}{c}{\emph{carry tortilla}}\\[0.5ex]
      \includegraphics[width=0.1\textwidth]{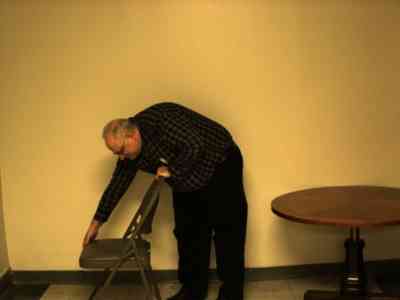}&
      \includegraphics[width=0.1\textwidth]{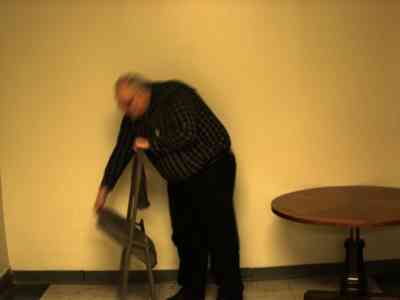}&
      \includegraphics[width=0.1\textwidth]{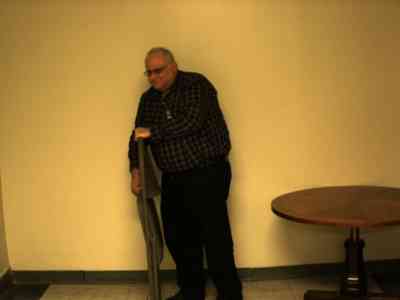}&
      \includegraphics[width=0.1\textwidth]{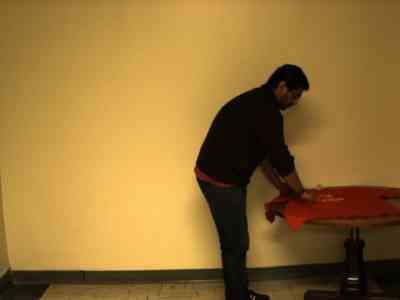}&
      \includegraphics[width=0.1\textwidth]{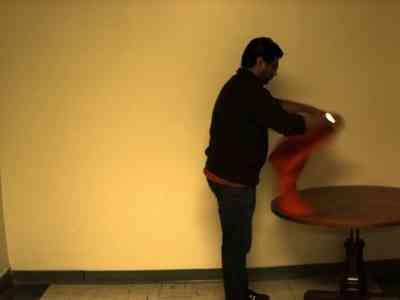}&
      \includegraphics[width=0.1\textwidth]{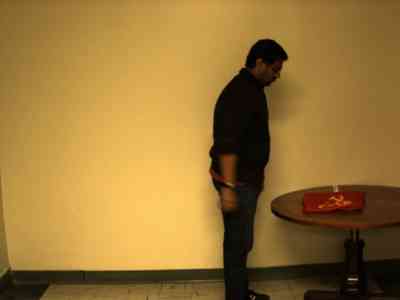}&
      \includegraphics[width=0.1\textwidth]{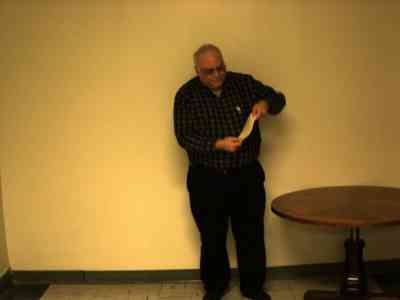}&
      \includegraphics[width=0.1\textwidth]{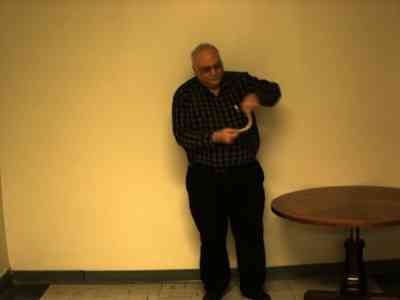}&
      \includegraphics[width=0.1\textwidth]{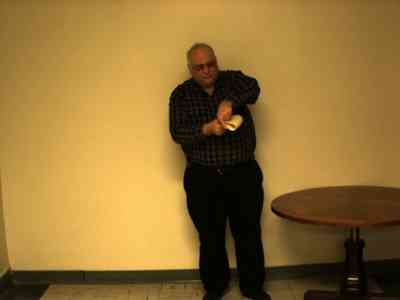}\\[-0.8ex]
      \multicolumn{3}{c@{\hspace{10pt}}}{\emph{fold chair}}&
      \multicolumn{3}{c@{\hspace{10pt}}}{\emph{fold shirt}}&
      \multicolumn{3}{c}{\emph{fold tortilla}}\\[0.5ex]
      \includegraphics[width=0.1\textwidth]{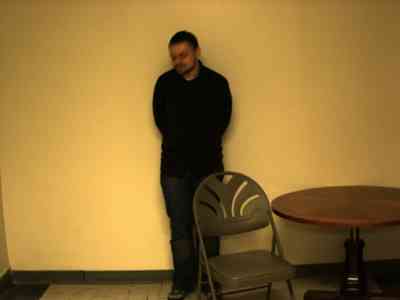}&
      \includegraphics[width=0.1\textwidth]{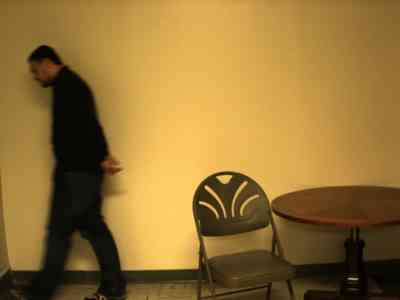}&
      \includegraphics[width=0.1\textwidth]{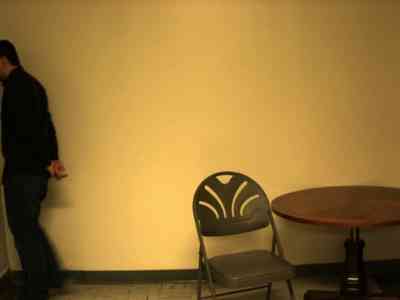}&
      \includegraphics[width=0.1\textwidth]{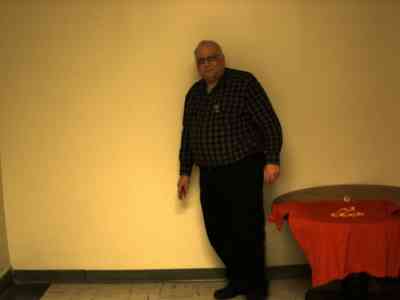}&
      \includegraphics[width=0.1\textwidth]{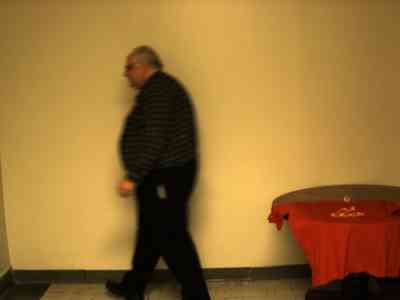}&
      \includegraphics[width=0.1\textwidth]{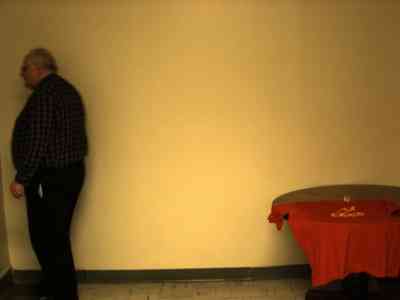}&
      \includegraphics[width=0.1\textwidth]{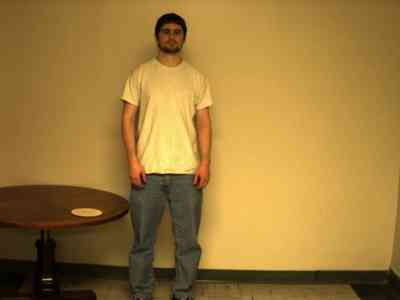}&
      \includegraphics[width=0.1\textwidth]{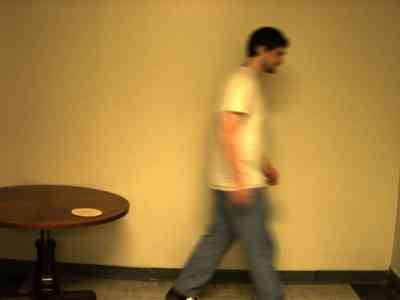}&
      \includegraphics[width=0.1\textwidth]{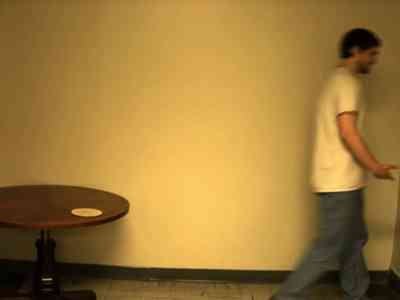}\\[-0.8ex]
      \multicolumn{3}{c@{\hspace{10pt}}}{\emph{leave chair}}&
      \multicolumn{3}{c@{\hspace{10pt}}}{\emph{leave shirt}}&
      \multicolumn{3}{c}{\emph{leave tortilla}}
    \end{tabular}
  \end{center}
  \caption{Key frames from sample stimuli in Experiment~2.
    Example stimulus videos are included in the supplementary material.}
  \label{fig:9events}
\end{figure}

We conducted a further experiment to evaluate the ability to recover
compositional semantics for entire sentences.
Subjects were shown videos that depict sentences of the form: the \emph{actor}
\emph{verb} the \emph{object} \emph{direction/location}.
They were asked to think about the sentence depicted in each video and
otherwise had no task.
Videos depicting three verbs (\emph{carry}, \emph{fold}, and \emph{leave}),
each performed with three objects (\emph{chair}, \emph{shirt}, and
\emph{tortilla}), each performed by four human actors, and each performed on
either side of the field of view were filmed for this task.
The verbs were chosen to be discriminable based on features described by
\citet{kemmerer2008}:
\begin{quote}
  \begin{tabular}{lrr}
    \emph{leave} & $-$state-change & $-$contact\\
    \emph{fold}  & $+$state-change & $+$contact\\
    \emph{carry} & $-$state-change & $+$contact\\
  \end{tabular}
\end{quote}
\noindent
Nouns were chosen based on categories found to be easily discriminable
by \citet{just2010}: \emph{chair} (furniture), \emph{shirt} (clothing), and
\emph{tortilla} (food) and also selected to allow each verb to be performed
with each noun.
Because these stimuli are not as ambiguous as the ones from Experiment~1,
they were not shown in pairs.
All stimuli enactments were filmed against the same nonvarying background,
which contained no other objects except for a table (Fig~\ref{fig:9events}).

This experiment, like Experiment~1, also used a rapid event-related
design.
We collected multiple videos, between 4 and 7, for each cross product of the
verb, object and human actor.
Variation along the side of field of view and direction of motion was
accomplished by mirroring the videos about the vertical axis.
Such mirroring induces variation in direction of motion (leftward
\vs\ rightward) for the verbs \emph{carry} and \emph{leave} and induces
variation in the location in the field of view where the verb \emph{fold} occurs
(left half \vs\ right half).
We presented 2s video clips at 10fps followed by an average of 4s (minimum 2s)
fixation.
Each run comprised 72 stimulus presentations spanning 244 captured brain
volumes, with eight runs per subject, and ended with 24s of fixation.
Each run was individually counterbalanced for each of the four conditions
(verb, object, actor, and mirroring).
We collected data for three subjects at Purdue University but discarded the
data for one of the three due to subject motion.
One subject did eight runs without exiting the scanner.
One subject exited the scanner between runs six and seven, which required
cross-session registration.
All subjects were aware of the experiment design, were informed of the intended
depiction of each stimulus prior to the scan, and were instructed to think of
the intended depiction after each presentation.

\begin{figure}
  \begin{center}
    \begin{tabular}{rccc}
      \raisebox{8ex}{(a)}&\multicolumn{3}{c}{\begin{tabular}[b]{ccccccc}
          \toprule
          &\multicolumn{6}{c}{\textbf{Classification Accuracy}}\\
      \cmidrule(l){2-7}
      \textbf{Subject}
      &\textbf{event}&\textbf{verb}&\textbf{object}&\textbf{actor}&
      \textbf{direction}&\textbf{location}\\
      \cmidrule{1-1}
      \cmidrule(l){2-7}
      1&55.38\%&82.12\%&64.41\%&32.81\%&83.33\%&68.23\%\\
      2&40.97\%&70.31\%&56.42\%&32.81\%&71.70\%&64.58\%\\[0.4ex]
      chance&11.11\%&33.33\%&33.33\%&25.00\%&50.00\%&50.00\%\\\bottomrule
    \end{tabular}}\\[1ex]
      \raisebox{-12pt}{(b)}&
      \includegraphics[width=0.25\textwidth]{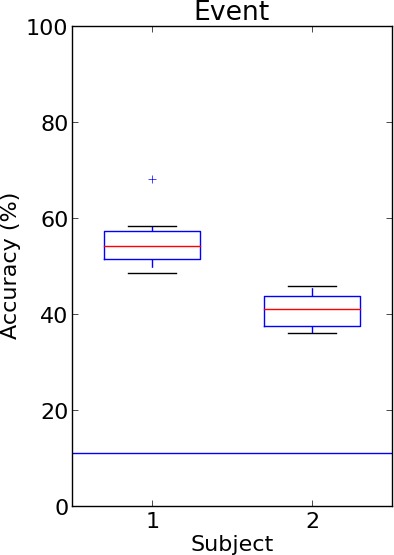}&
      \includegraphics[width=0.25\textwidth]{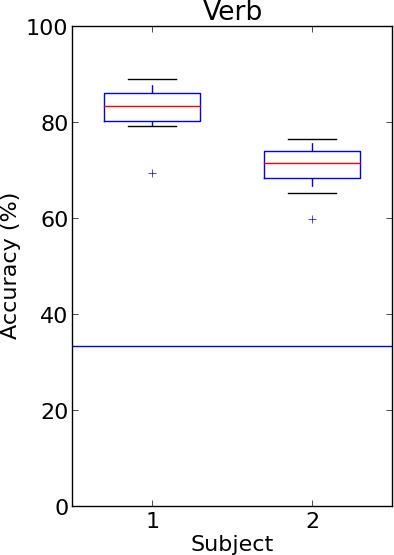}&
      \includegraphics[width=0.25\textwidth]{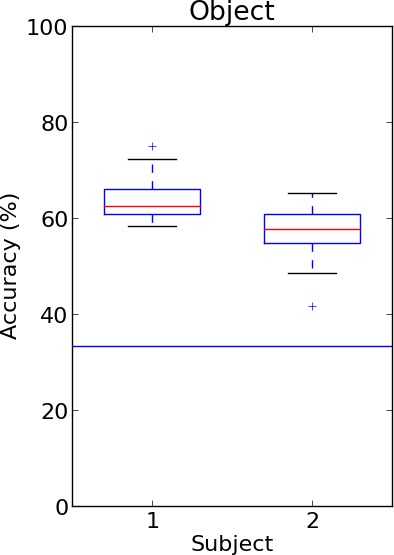}\\
      &\includegraphics[width=0.25\textwidth]{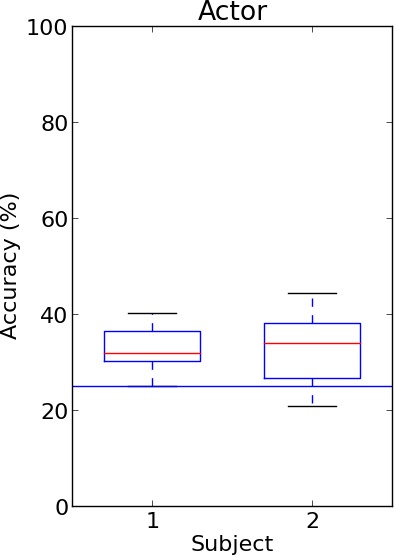}&
      \includegraphics[width=0.25\textwidth]{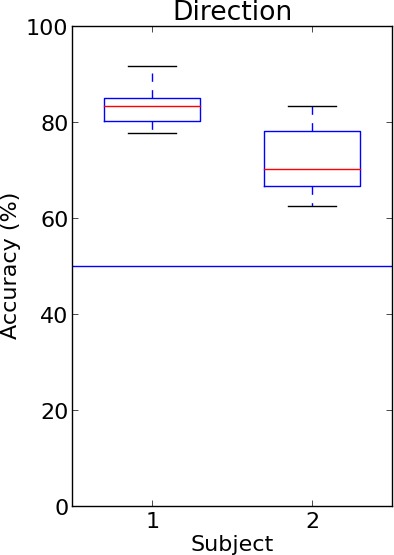}&
      \includegraphics[width=0.25\textwidth]{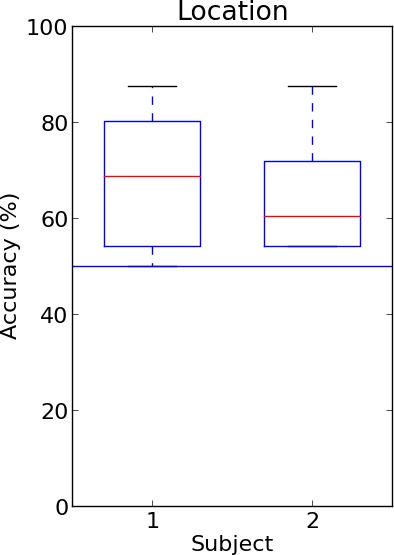}\\

      \raisebox{-15pt}{(c)}&
      \includegraphics[width=0.25\textwidth]{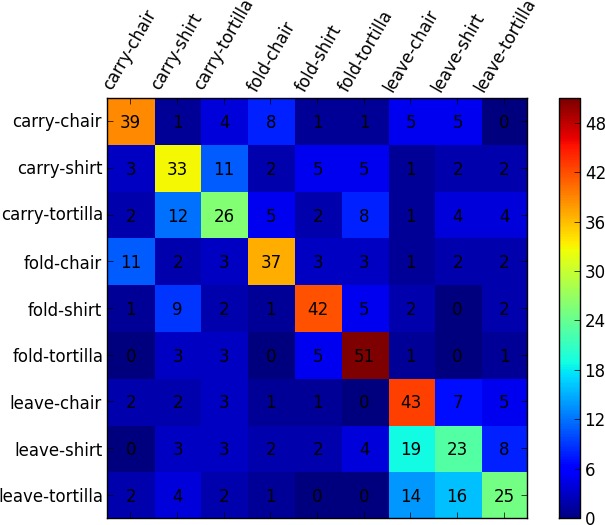}&
      \includegraphics[width=0.25\textwidth]{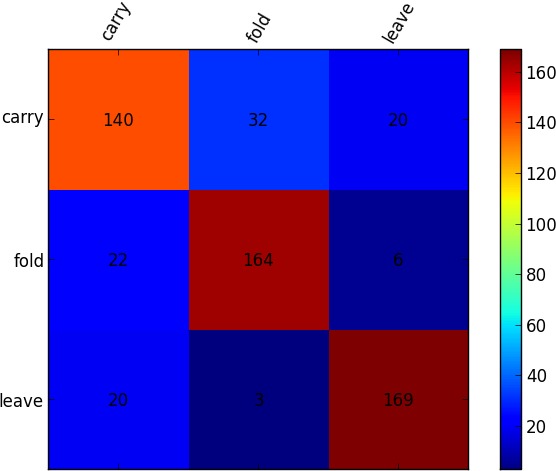}&
      \includegraphics[width=0.25\textwidth]{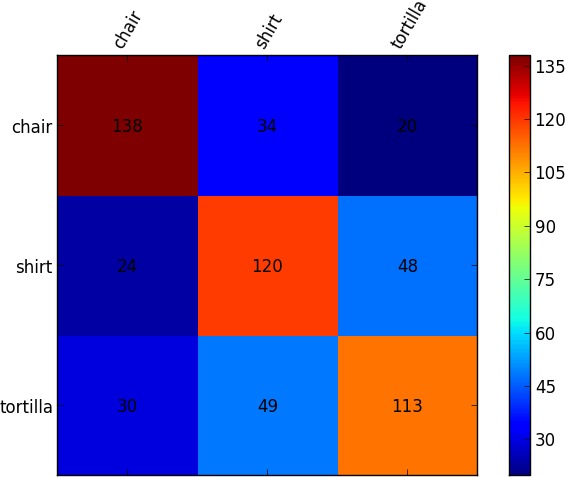}\\
      &\textbf{event}&\textbf{verb}&\textbf{object}\\
      &\includegraphics[width=0.25\textwidth]{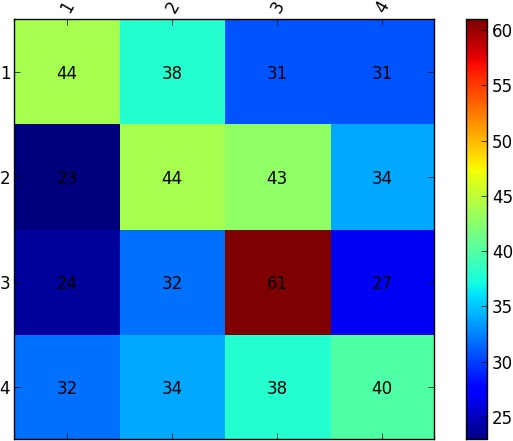}&
      \includegraphics[width=0.25\textwidth]{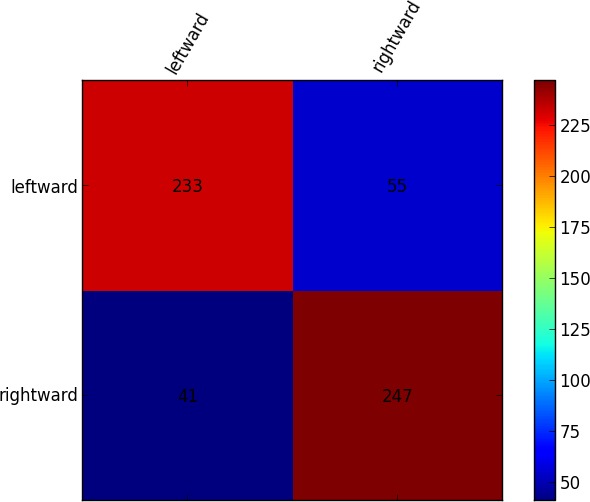}&
      \includegraphics[width=0.25\textwidth]{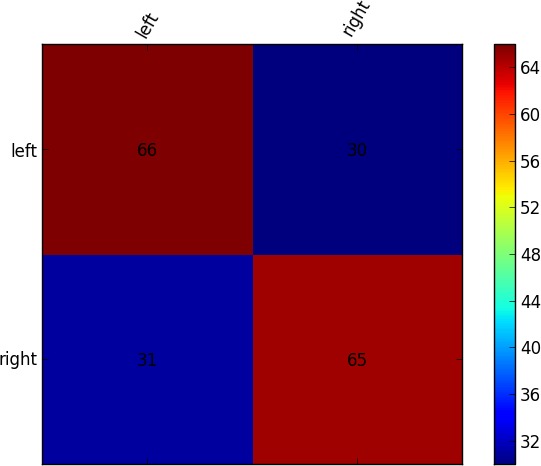}\\
      &\textbf{actor}&\textbf{direction}&\textbf{location}
    \end{tabular}
  \end{center}
  \caption{Results for Experiment~2.
    (a)~Per-subject mean classification accuracies averaged across fold.
    Note that all six analyses perform above chance.
    (b)~Per-subject classification accuracies showing the means and variances
    of performance across the different folds for each class.
    The horizontal line indicates chance performance.
    (c)~Corresponding confusion matrices, averaged across subject and fold.
    Note that they are mostly diagonal.}
  \label{fig:9events-results}
\end{figure}

This experimental design supports the following classification analyses:
\begin{compactdesc}
\item[event] one-out-of-9 verb\&noun (\emph{carry}, \emph{fold}, and
  \emph{leave}, each performed on \emph{chair}, \emph{shirt}, and
  \emph{tortilla})
\item[verb] one-out-of-3 verb (\emph{carry}, \emph{fold}, and \emph{leave})
\item[object] one-out-of-3 noun (\emph{chair}, \emph{shirt}, and
  \emph{tortilla})
\item[actor] one-out-of-4 actor identity
\item[direction] one-out-of-2 motion direction for \emph{carry} and \emph{leave}
  (leftward \vs\ rightward)
\item[location] one-out-of-2 location in the field of view for \emph{fold}
  (right \vs\ left)
\end{compactdesc}
The analysis performed was exactly the same as that for Experiment~1,
including eight-fold cross validation for each of our analyses, where runs
constituted folds.
Fig.~\ref{fig:9events-results} presents an overview of the results along with
per-subject classification accuracies and aggregate confusion matrices for the
each of the above analyses.
Note that we achieve significantly above-chance performance on all six analyses
with only a single fold for a single subject across all six analyses
performing below chance.

\textbf{Verb} performance is well above chance (76.22\%, chance 11.11\%).
This replicates Experiment~1 with different videos and a new verb and adds to
the evidence that brain activity corresponding to verbs can reliably be decoded
from fMRI scans.
\textbf{Object} performance was significant as well (60.42\%, chance 33.33\%).
Given neural activation, we can decode which object the subjects are thinking
about.
We know of no other work that decodes brain activity corresponding to objects
from videos.
The fact that the verb and object can be decoded independently already provides
evidence of argument compositionality.
Were the neural representations not compositional at this level, decoding would
not be possible.
For example, if the representation of \emph{carry} was neurally encoded as a
combination of \emph{walk} and a particular object, verb performance would not
exceed chance, because our experiment is counterbalanced with respect to the
object with which the action is being performed.
While this indicates that the representations for verbs and objects are
independent of each other to some degree, we also seek to quantify the level of
independence.
If the representation of \emph{carry} is somewhat different depending on which
object is being carried, we expect that performance would increase when we
jointly classify the object and the verb.
This seems to not be the case.
The accuracy of \textbf{event} is almost identical to the joint independent
accuracy of \textbf{verb} and \textbf{object}:
$\text{0.5538}\approx\text{0.5289}=\text{0.8212}\times\text{0.6441}$
(subject~1) and
$\text{0.4097}\approx\text{0.3967}=\text{0.7031}\times\text{0.5642}$
(subject~2), indicating that the representation of these verbs is independent
of the objects that the verbs are being performed with.
This is also confirmed by the confusion matrix for \textbf{event} in
Fig.~\ref{fig:9events-results}(c) which remains diagonal.

To decode complex brain activity corresponding to an entire sentence, we can
combine \textbf{actor}, \textbf{verb}, \textbf{object}, and \textbf{direction}
or \textbf{location}.
We perform significantly above chance on this one-out-of-72
(4$\times$3$\times$3$\times$2) classification:
\begin{equation*}
  \begin{array}{@{}lr@{}}
\text{0.3281}\times\text{0.8212}\times\text{0.6441}\times(\frac{\text{1}}{\text{3}}\times\text{0.6823}+\frac{\text{2}}{\text{3}}\times\text{0.8333})=\textbf{0.1359}\gg\textbf{0.0139}=\frac{\text{1}}{\text{72}}
&\text{(subject~1)}\\[0.5ex]
\text{0.3281}\times\text{0.7031}\times\text{0.5642}\times(\frac{\text{1}}{\text{3}}\times\text{0.6458}+\frac{\text{2}}{\text{3}}\times\text{0.7170})=\textbf{0.0902}\gg\textbf{0.0139}=\frac{\text{1}}{\text{72}}&\text{(subject~2)}
  \end{array}
\end{equation*}
(Since \textbf{direction} applied to \emph{carry} and \emph{leave} while
\textbf{location} disjointly applied to \emph{fold}, this yields a binary
classification task across all verbs.)
Thus we are able to classify \textbf{\emph{entire sentences}} compositionally
from their individual words.

\begin{figure}[t]
  \begin{center}
    \includegraphics[width=\textwidth]{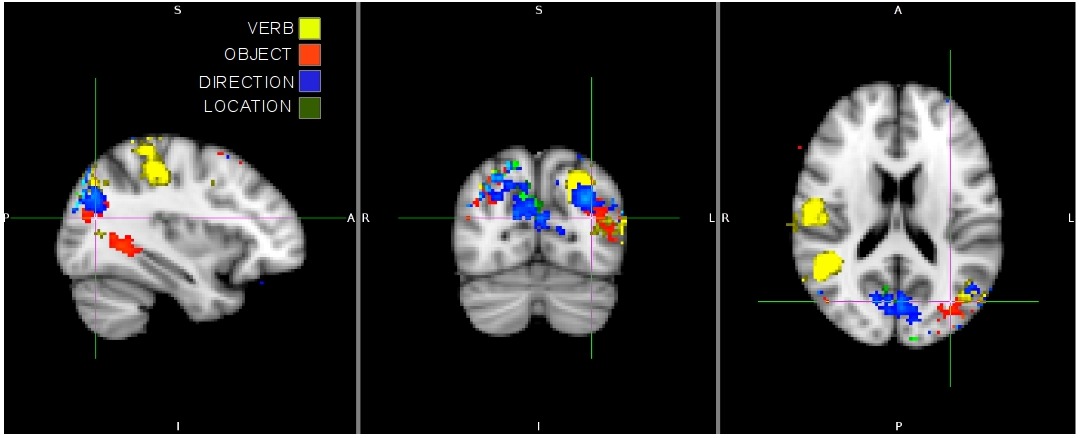}
  \end{center}
  \caption{Searchlight analysis for Experiment~2 indicating the classification
    accuracy of different brain regions on the anatomical scans from subject~1
    averaged across stimulus, class, and run.}
  \label{fig:9events-searchlight}
\end{figure}

To locate regions of the brain used in the previous analyses, we applied the
same searchlight linear-SVM method that was performed in Experiment~1 to
subject~1's data from this experiment and identified similar areas in
visual-pathway, parietal, and prefrontal areas.
The resulting ROIs, shown in Fig.~\ref{fig:9events-searchlight}, are overlaid
and color coded according to the specific visual feature being decoded.
In general, it is clear that the decoding is sensitive to action/category
information and various visual object-and-motion features.
Many of the same regions active for \textbf{verb} in Experiment~1 also
show activity in this experiment.
\textbf{Direction} and \textbf{location} activity is present in the visual
cortex with significant \textbf{location} activity occurring in the early
visual cortex.
\textbf{Object} activity is present in the temporal cortex, and agrees with
previous work on object-category encoding \citep{Gazzaniga2008}.

\section{Conclusion}
\label{sec:conclusion}

We have demonstrated that it is possible to read a subject's brain activity and
decode a complex action tableau corresponding to a sentence from its
constituents.
To do so, we showed novel work which decodes brain activity associated with
verbs and simultaneously recovers lexical aspects of different parts of speech.
Our results indicate that the neural representations for verbs and objects
compose together to form the meaning of a sentence apparently without modifying
one another.
These results indicate that representations which attempt to decompose meaning
into constituents may have a neural basis.

\subsubsection*{Acknowledgments}

AB, NS, and JMS were supported, in part, by Army Research Laboratory (ARL)
Cooperative Agreement W911NF-10-2-0060.
CX and JJC were supported, in part, by ARL Cooperative Agreement
W911NF-10-2-0062 and NSF CAREER grant IIS-0845282.
CDF was supported, in part, by NSF grant CNS-0855157.
CH and SJH were supported, in part, by the McDonnell Foundation.
BAP was supported, in part, by Science Foundation Ireland grant 09/IN.1/I2637.
The views and conclusions contained in this document are those of the authors
and should not be interpreted as representing the official policies, either
express or implied, of the supporting institutions.
The U.S. Government is authorized to reproduce and distribute reprints for
Government purposes, notwithstanding any copyright notation herein.
Dr.\ Gregory G. Tamer, Jr.\ provided assistance with imaging and analysis.

\bibliographystyle{abbrvnat}
\bibliography{arxiv2013d}
\end{document}